\documentclass[twocolumn,english,pra,showpacs,nofootinbib,superscriptaddress]{revtex4}
\usepackage{graphicx}
\usepackage{epsfig}
\usepackage{subfigure}
\usepackage{mathrsfs}
\usepackage{epstopdf}
\usepackage{amsmath}

\usepackage{amssymb}

\usepackage{setspace} 

\usepackage[normalem]{ulem}
\usepackage{color}



\begin{document}
\title{Finite-Temperature Study of Bose-Fermi Superfluid Mixtures}
\author{B. Ramachandhran}\affiliation{Department of Physics \&
  Astronomy, and Rice Quantum Institute, Rice University, Houston, TX
  77005, USA} \author{S. G. Bhongale} \affiliation{Department of
  Physics \& Astronomy, and Rice Quantum Institute, Rice University,
  Houston, TX 77005, USA} \affiliation{Department of Physics \&
  Astronomy, George Mason University, MS 3F3, Fairfax, VA 22030}
\author{H. Pu}\affiliation{Department of Physics \& Astronomy, and
  Rice Quantum Institute, Rice University, Houston, TX 77005, USA}
  
  \begin{abstract}   
    Ultra-cold atom experiments offer the unique opportunity to study
    mixing of different types of superfluid states. Our interest is in
    superfluid mixtures comprising particles with different statistics-- Bose and Fermi.
    Such scenarios occur naturally, for example, in
    dense QCD matter. Interestingly, cold atomic
    experiments are performed in traps with finite spatial extent,
    thus critically destabilizing the occurrence of various
    homogeneous phases. Critical to this analysis
    is the understanding that the trapped system can undergo phase
    separation, resulting in a unique situation where phase transition
    in either species (bosons or fermions) can overlap with the phase
    separation between possible phases. In
    the present work, we illustrate how this
    intriguing interplay manifests in an interacting 2-species atomic
    mixture -- one bosonic and another fermionic with two spin
    components -- within a realistic trap configuration. We further
    show that such interplay of transitions can render the nature of
    the ground state to be highly sensitive to the experimental
    parameters and the dimensionality of the system.
    \end{abstract}
\pacs{ 67.85.Pq, 67.85.-d, 67.85.Lm}\date{\today}
\maketitle
\section{Introduction}
\label{secintro}
Ultra-cold trapped-atom experiments offer the
unique possibility to understand many-body physics beyond what can be
explored in typical condensed matter settings
\cite{BlochSimulate}. Essentially, they provide clean many-body
systems in which attributes like density, dimensionality and
interactions, may be controlled with commendable precision
\cite{reviewBEC, reviewFermi, FeshbachPure, opticalfeshbach}.  As a
result, from a theoretical perspective, there are broadly two kinds of
challenges: (1) investigate configurations appropriate for emulating
many-body theory models, thereby allowing for a systematic
verification of claims made in the condensed matter context, and (2)
investigate new configurations extremely difficult to realize in
material settings. While the former program has 
proved quite successful with demonstrations of, for example, Mott
insulator to superfluid transition with ultra-cold $^{87}$Rb atoms in
an optical lattice \cite{MottGreiner}, the latter is just beginning to
attract attention with several new experiments comprising degenerate
mixtures of bosons and fermions, of same or different species
being set up \cite{KetterleSympathetic, FeshbachMix}. A
potentially rich scenario in this context is provided by an
atomic mixture comprising superfluids of two kinds- bosonic and
fermionic. Studying this system may also have strong implications for,
say nuclear physics, as a recent proposal
investigates the intriguing possibility of
simulating dense QCD matter with superfluid atomic mixtures
\cite{QCD}.  Further, considering that in condensed matter setting,
the analogous $^3$He-$^4$He superfluid mixture is difficult to realize
\cite{Helium}, achieving Bose-Fermi superfluid mixtures with
ultra-cold atoms maybe an important step towards understanding
corresponding occurrences in a broader context.

In analyzing experiments with ultra-cold Bose-Fermi mixtures, it is
important to understand the effects of inhomogeneity due to
traps. These effects are at the heart of determining the
stability of possible thermodynamic
phases in a given experiment. To this end, we construct the
finite-temperature phase diagram of an interacting 3-dimensional (3-d)
mixture comprising of two fermions (spin $\uparrow$ \& $\downarrow$)
of one species and a bosons of another. To draw
such a phase diagram, it is important to
understand the interplay between the following two phenomena: (1)
phase transition that occurs near a critical temperature where
suddenly an order parameter corresponding to one of the species
nucleates. In fact, the critical temperature of such a transition may
itself depend intricately on the state of the second
species. Moreover, the already nucleated phase may subsequently be
drastically affected in a certain region of trap due to the nucleation
of a new phase, corresponding to the second species, as the system is
further cooled and crosses below a lower critical temperature. (2)
phase separation between possible phases, a
phenomenon unique to trapped configurations. It also implies that the
trap potential can simultaneously accommodate one or more of the
phases as determined by the experimental parameters. Thus, remarkably
what phase/phases will be observed will critically depend on the trap
geometry. This, in fact is a very important observation implying the
possibility of tuning the trap parameters such that a desired density
profile is observed only if a certain phase has nucleated. On top of
all this, the dimensionality of the trapped system, whether we
consider a 3-d or a 1-d trap, will also largely determine what phase
is energetically favorable for phase separation. 

While various possibilities discussed above exist and
some insight may be borrowed from previous studies on pure Bose and
Fermi superfluids, the intrinsically new nature of Bose-Fermi
superfluid mixtures strongly motivates us to derive a framework within
which an elaborate finite temperature phase diagram can be
generated. Also, such finite temperature studies comprising
interacting fermions have never been performed in
the past. The paper is organized as follows. In Sec.~\ref{secTheory}, we first
review the theory for analyzing the thermodynamic instabilities of the
Bose-Fermi mixture. While the technique is quite standard and maybe
found elsewhere, to our knowledge this is the first instance where it
has been applied for deriving the finite temperature phase diagram of
the inhomogeneous mixture comprising of bosons and fermions, both in
the superfluid phase. As discussed in the previous paragraphs, the
trap introduces multiple scenarios that are new to these systems
making the analysis complicated. Therefore as a warmup, in
Sec.~\ref{secZeroT} we illustrate our method by first considering the
simplest case of the $T=0$ superfluid mixture in 3-d. The finite
temperature phase diagram for the 3-d Bose Fermi mixture will be
derived in Sec.~\ref{secFiniteT}. Finally, in Sec.~\ref{secLDA}, we
will discuss the implication of the phase diagram for a \emph{trapped}
Bose-Fermi mixture by introducing a spatially varying chemical
potential in the spirit of a Local Density Approximation (LDA),
followed by brief discussion of the dependance on dimensionality in
Sec.~\ref{sec1D}.
\section{Theory}
\label{secTheory}
We begin by writing the Hamiltonian for
the interacting Bose-Fermi mixture in the form
\begin{equation}
\hat{H}= \underbrace{\hat{H}_{b}  - \mu_b \,  \hat{N}_b}_{\hat{\mathcal{H}}_b} + \underbrace{\hat{H}_{f}  - \mu_f \,  \hat{N}_f  + U_{bf} \, \hat{N}_b \, \hat{N}_f}_{\hat{\mathcal{H}}_{bf}},
\label{Hamtot}
\end{equation}
where the subscript $b$ ($f$) stands for bosons (fermions), $\mu$'s
represent corresponding chemical potentials, $\hat{N}$'s the
corresponding number operators and $U_{bf}$ denotes the interaction
energy between bosons and fermions. Our interest is in studying this
interacting Bose-Fermi mixture in the vicinity of the superfluid
critical temperature $T_c$ of the fermions. Of course, it is true that
the $T_c$ itself will be modified due to the presence of Bose
component. Further, the phase of fermions may modify the critical
temperature for the condensation of the Bose component,
$T_{BEC}$. Thus, while the general problem is indeed complicated, we
focus our attention on the situation when $T_{BEC}$ is much greater
than $T_c$, typically the case in most trapped experiments
\cite{Onofrio}. This allows us to work in the Thomas-Fermi limit of
the Bose component by neglecting its kinetic energy. We represent the
contact interaction strength between a pair of bosons as $\lambda_b =
U_b \,V = 4 \pi \hbar^2 a_b/m$, where $U_b$ is the interaction energy
of bosons, $V$ is the volume, $a_b$ is the $s$-wave boson-boson
scattering length, assumed to be positive implying repulsive
interactions, and $m$ is the mass of bosonic atom. For large boson
number $N_b$, total pairs of bosons is approximately $N_b^2/2$ and
hence, $\hat{H}_b$ is simply a constant given by $U_b N_b^2 /2$. Thus,
the contribution to the free energy density arising from just the
bosonic component is
\begin{equation}
f_b =  \langle \hat{\mathcal{H}}_b \rangle = \lambda_b \, \frac{n_b^2}{2}  -\mu_b \, n_b, \label{freeb} 
\end{equation}
where $n_b = N_b/V$ is the boson density.

Now, we focus on the $\hat{\mathcal{H}}_{bf}$ part of the Hamiltonian
and write it explicitly in second quantized form as
\begin{widetext}
\begin{equation}
\hat{\mathcal{H}}_{bf} = \sum_{k,\sigma} ( \varepsilon_k - \mu_f) \,c^{\dagger}_{k,\sigma} c_{k,\sigma} +  \lambda_{f} \sum_{k,k',q} c^{\dagger}_{k+q,\uparrow} \,c^{\dagger}_{-k,\downarrow} c_{-k'+q,\downarrow} \,c_{k',\uparrow} + \lambda_{bf} \, n_b \sum_{k,\sigma} \,c^{\dagger}_{k,\sigma} c_{k,\sigma}.
\label{Hamfbf}
\end{equation}
\end{widetext}
Here $\varepsilon_k = \hbar^2 k^2 / 2 m$ and
$c^{\dagger}_{k,\sigma}$($c_{k,\sigma}$) is the creation
(annihilation) operator for a fermion with momentum $k$ and spin
$\sigma$. Further, the boson-fermion interaction, which is
typically short range, is described by a
$\delta$-potential contact interaction with strength given by
$\lambda_{bf} = 2 \pi \hbar^2 a_{bf}/\mu_m$, where $a_{bf}$ is the
corresponding $s$-wave scattering length and $\mu_m$ is the reduced
mass of the boson-fermion system. Here we will confine
our analysis to the repulsive regime with $a_{bf}>0$. Similarly, we
describe the fermion-fermion interaction by the contact interaction
strength $\lambda_{f} = 4 \pi \hbar^2 a_{f}/m$, where $a_f$ is the
corresponding $s$-wave scattering length. Here, since the interaction
is $s$-wave, only unequal-spin fermions interact. Also, we are
interested in the superfluid regime, which occurs for attractive
interactions, thus we assume $a_f<0$.  Now, in the superfluid state
with BCS-type pairing \cite{Grimm, Servaas}, the center-of-mass
momentum, $q$, of the Cooper pair is set to
zero allowing $\hat{\mathcal{H}}_{bf}$ to
be simply 
\begin{equation}
\hat{\mathcal{H}}_{bf} =  \sum_{k,\sigma} \xi_k \,c^{\dagger}_{k,\sigma} c_{k,\sigma} - |\lambda_{f}| \, \sum_{k,k'} c^{\dagger}_{k,\uparrow} \,c^{\dagger}_{-k,\downarrow} c_{-k',\downarrow} \,c_{k',\uparrow},
\label{Hamfbf1}
\end{equation}
with $\xi_k = \varepsilon_k - \mu_f + \lambda_{bf} n_b$. One can
  immediately notice that this is just the usual BCS Hamiltonian with
  a modified chemical potential, hence can be diagonalized with the
  usual Bogoliubov transformation \cite{Altland}. Firstly, defining the mean-field order parameter $\Delta=|\lambda_{f}| \sum_{k'} \langle c_{-k',\downarrow} \,c_{k',\uparrow} \rangle$ and its complex conjugate ${\Delta}^*$, we write $\hat{\mathcal{H}}_{bf}$ as
\begin{widetext}
\begin{eqnarray}
\hat{\mathcal{H}}_{bf} \stackrel{M.F}{=}  \sum_{k} \xi_k \big(\,c^{\dagger}_{k,\uparrow} c_{k,\uparrow} + c^{\dagger}_{k,\downarrow} c_{k,\downarrow} \big)
+ \frac{\,|\Delta|^2}{|\lambda_{f}|} 
-\big( {\Delta}^* \sum_{k'} c_{-k',\downarrow} \,c_{k',\uparrow} + \Delta \sum_{k} c^{\dagger}_{k,\uparrow} \,c^{\dagger}_{-k,\downarrow} \big).  
\label{Hamfbf2}
\end{eqnarray}
\end{widetext}
Re-writing the above in terms of the \emph{Nambu spinor} $\Psi_k^{\dagger} = ( c^{\dagger}_{k,\uparrow} \,, c_{-k,\downarrow} )$ and its hermitian conjugate $\Psi_k$, we have
\begin{equation}
\hat{\mathcal{H}}_{bf} = \sum_{k} \Psi_k^{\dagger} \left( \begin{array}{cc} \xi_k & - \Delta \\  - {\Delta}^* & - \xi_k \end{array} \right) \Psi_k + \sum_{k} \xi_k + \frac{|\Delta|^2}{|\lambda_{f}|}\,.
\label{ham1}
\end{equation}
Now the Bogoliubov transformation immediately gives
\begin{eqnarray}
\hat{\mathcal{H}}_{bf} &=& \sum_{k} \left( \alpha^{\dagger}_{k,\uparrow} \,, \alpha_{-k,\downarrow} \right) \left( \begin{array}{cc} E^{+}_k & 0 \\  0 & E^{-}_k \end{array} \right) \left(  \begin{array}{cc} \alpha_{k,\uparrow} \\ \alpha^{\dagger}_{-k,\downarrow} \end{array} \right) \nonumber \\
&+& \sum_{k} \xi_k + \frac{\Delta^2}{|\lambda_{f}|},
\label{ham2}
\end{eqnarray}
where the eigenenergies are $E_k^{\pm} = \pm \sqrt{\xi_k^2 + \Delta^2}$ and $\Delta$ is assumed real \cite{Altland1}. The operator $\alpha^{\dagger}_{k,\uparrow}$ ($\alpha_{k,\uparrow}$) creates (annihilates) Bogoliubov quasiparticles that are distributed according to the Fermi-Dirac distribution $f_k = 1/(1+e^{\beta E_k})$ with $\beta =
1/k_B \,T$. Thus the relevant thermodynamic potential is given by
\begin{equation}
\langle \hat{\mathcal{H}}_{bf} \rangle - T S 
  = \sum_{k} (\xi_k - E_k)  + \frac{\Delta^2}{|\lambda_{f}|}  - \frac{2}{\beta} \sum_k  \textrm{ln}\, (1+e^{-\beta \, E_k})
\label{freefbf}
\end{equation}
where $S$ is the entropy. The derivations and mean-field analysis presented henceforth is quantitatively exact only when the interactions are weak. Our analysis is only qualitatively correct in the strong interaction limit, where a strong-coupling theory presented along the lines of Ref.~\cite{DawWei} would be quantitatively more accurate.
\subsection{Free energy, Equilibrium and dynamical stability conditions}
\label{derivefree}
Free energy density of the interacting mixture comprising of bosons and
fermions, both in the superfluid state, can now be
written from Eqs.~(\ref{freeb})
and (\ref{freefbf}):
\begin{equation*}
f = \frac{\lambda_b n_b^2}{2}  -\mu_b \, n_b + \sum_{k} (\xi_k - E_k)  + \frac{\Delta^2}{|\lambda_{f}|}  - \frac{2}{\beta}\sum_k  \textrm{ln}\, (1+e^{-\beta \, E_k}).
\end{equation*}
As one can immediately notice, $f$ depends on
numerous parameters:
interaction strengths $\lambda_{\{b,f,bf\}}$ (in-turn, the scattering
lengths $a_{\{b,f,bf\}}$), particle densities $n_{\{b,f\}}$, chemical
potentials $\mu_{\{b,f\}}$, BCS superfluid order parameter $\Delta$
and temperature $T$. It is quite evident that the phase space of this
interacting mixture is huge and thus an exhaustive study is
impossible. However, noticing the fact that not all of these
parameters are independent, we adopt the following
scheme that was first introduced in Ref.~\cite{Satyan1Dp}, allowing us to investigate the experimentally relevant region of the phase space: (1) fix
parameters $\lambda_b$, $\lambda_{bf}$ and $\mu_f$ and perform our
analysis at fixed values of $T$ \cite{muf}; (2) we project the
multi-dimensional phase diagram in the $\{n_b,\Delta\}$ phase space;
(3) the remaining dependent parameters $\mu_b$, $n_f$
and $\lambda_f$ are determined by the equilibrium stability conditions
to be derived below.\\
\\
\emph{(1) First derivative conditions:} First of these is the Gap equation obtained as the extremum of  $f$ with respect to $\Delta$ and provides the self-consistent value of the interaction strength parameter $\lambda_{f}$:
\begin{equation*}
\frac{\partial f}{\partial \Delta} = 0 \implies \frac{1}{|\lambda_{f}|} =  \sum_k \frac{1}{2 \, E_k} \, \textrm{tanh} \, \big( \frac{\beta E_k}{2} \big)\,.
\end{equation*}
However, in three dimensions, the
  momentum sum in the above expression diverges, an artifact of the
  contact interaction approximation. This unphysical effect is easily eliminated by an
  appropriate regularizing prescription. One of the easiest and
  convenient methods is to subtract the diverging piece:
\begin{equation}
 \frac{1}{|\lambda_{f}|}  = \frac{m}{4 \pi \hbar^2 |a_{f}|}  = \sum_k \frac{1}{2 \, E_k} \textrm{tanh} \, \big( \frac{\beta E_k}{2} \big) - \frac{1}{2 \, \varepsilon_k}.
 \label{gap}
\end{equation}
Correspondingly, we upgrade the free energy
  $f$ to the regularized $f_{\text{reg}}$ such that the extremum condition automatically reproduces the
  regularized version of the Gap equation \cite{diverges}
\begin{eqnarray}
f_{\text{reg}} (n_b, \Delta) &=& \lambda_b \, \frac{n_b^2}{2}  -\mu_b \, n_b + \sum_{k} (\xi_k - E_k + \frac{\Delta^2}{2 \, \varepsilon_k})\nonumber\\&&
 + \frac{\Delta^2}{|\lambda_{f}|}  - \frac{2}{\beta} \sum_k  \textrm{ln}\, (1+e^{-\beta \, E_k})\,.
\label{freetot}
\end{eqnarray}
The above step is essential for our case since we will be
  eventually interested not only in the stability of the Fermi system
 but that of the combined Bose-Fermi system. The presence of
  bosons affects the self-consistent value of $\lambda_f$, through the
  combination $\lambda_{bf}\,n_b$.  Next we consider the
  variation with respect to the fermion chemical potential
  $\mu_f$. This produces the familiar equation determining the fermion
  number density $n_f$:
\begin{equation}
\frac{\partial (f_{\text{reg}}+\mu_f \, n_f)}{\partial \mu_f} = 0 \implies n_f = \sum_k 1 - \frac{\xi_k}{E_k} \, \textrm{tanh} \, \big( \frac{\beta E_k}{2} \big)\,.
\label{number}
\end{equation}
Finally, the last of dependent parameters, the boson chemical potential
  $\mu_b$ is determined by minimizing $f_{\text{reg}}$ with respect to
  the boson density. This leads to the modified Thomas-Fermi equation
  given by
\begin{equation}
\partial f_{\text{reg}} / \partial n_b = 0 \implies \mu_b = \lambda_b n_b + \lambda_{bf}  n_f.
\label{mub}
\end{equation}
\\
\emph{(2) Second derivative conditions:} The second derivatives at the
extremum points derived above provide the dynamical stability
criterion for the mixture via positive definiteness of the Hessian
matrix ${\cal M}$. The relevant Hessian matrix elements are:
\begin{widetext}
 \begin{eqnarray}
    {\cal M}_{11}&=&\frac{\partial ^2f_{\text{reg}}}{\partial n_{b}^{2}} = \lambda_b - \lambda^{2}_{bf}  \sum_k \frac{\Delta^{2}}{E^{3}_k} \, \textrm{tanh}(\frac{\beta E_k}{2} ) - \frac{\beta}{2} \frac{\xi^{2}_k}{E^{2}_k} \, \textrm{sech} ( \frac{\beta E_k}{2} )^2; \\
    {\cal M}_{22}&=&\frac{\partial ^2f_{\text{reg}}}{\partial \Delta^2} = \sum_k \frac{\Delta^2}{E^{3}_k} \, \textrm{tanh} ( \frac{\beta E_k}{2} ) - \frac{\beta}{2}  \frac{\Delta^2}{E^{2}_k} \, \textrm{sech} ( \frac{\beta E_k}{2} )^2;\\
    {\cal M}_{12}={\cal M}_{21}&=&\frac{\partial ^2f_{\text{reg}}}{\partial \Delta
      \, \partial n_b}  = \lambda_{bf} \sum_k \frac{\Delta \, \xi_k}{E^{3}_k}
    \, \textrm{tanh} ( \frac{\beta E_k}{2} ) - \frac{\beta}{2}
    \frac{\Delta \, \xi_k}{E^{2}_k} \, \textrm{sech} ( \frac{\beta
      E_k}{2} )^2.
\label{Hall}
\end{eqnarray}
\end{widetext}
\section{Finite Temperature Phase Diagram}
\label{secPD}
\begin{figure*}[t]
  \centering
  \subfigure{\label{pdT0}\includegraphics[width=2.75in,height=2.5in,clip]{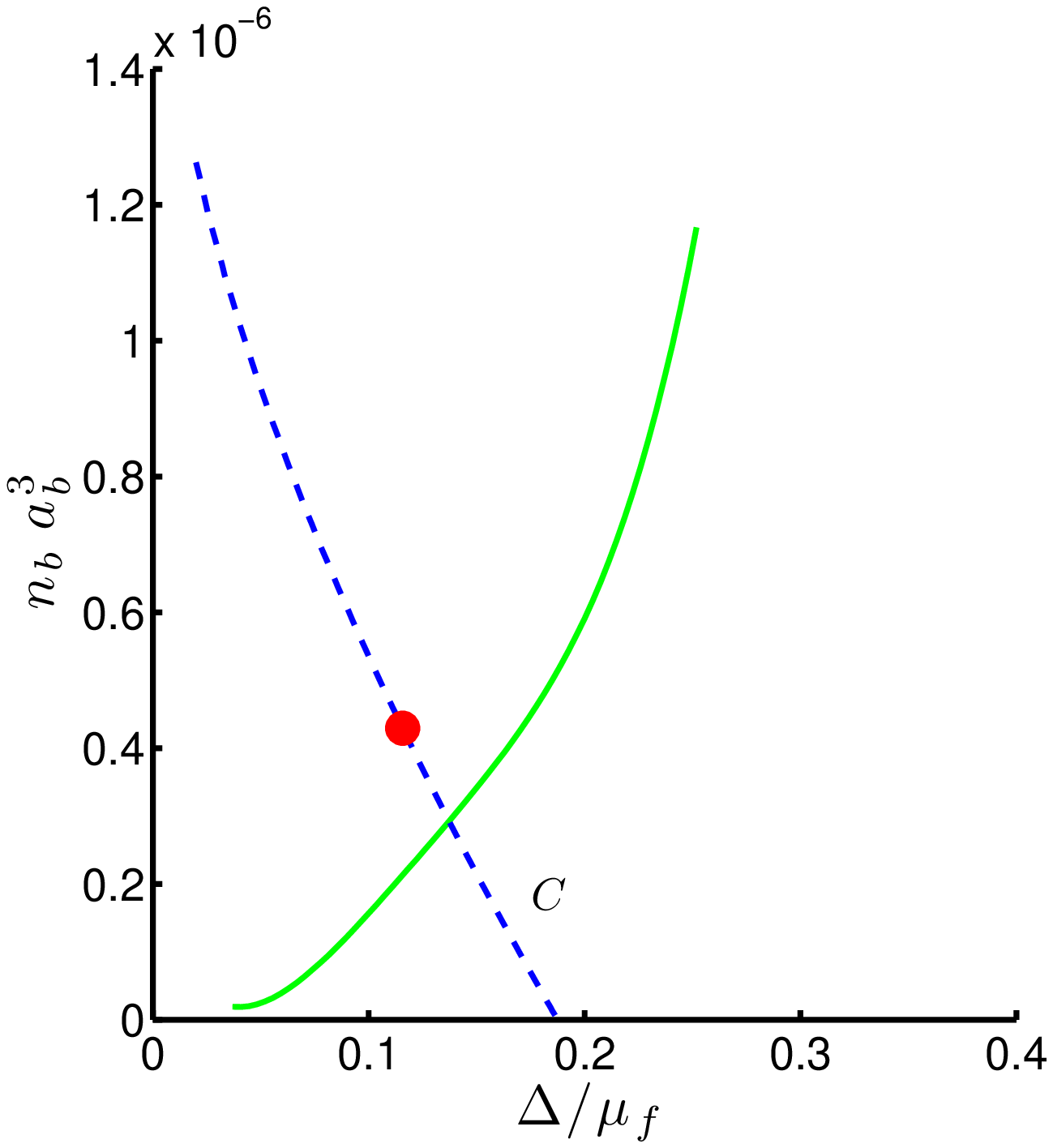}}
  \subfigure{\label{freeT0}\includegraphics[width=2.75in,height=2.5in,clip]{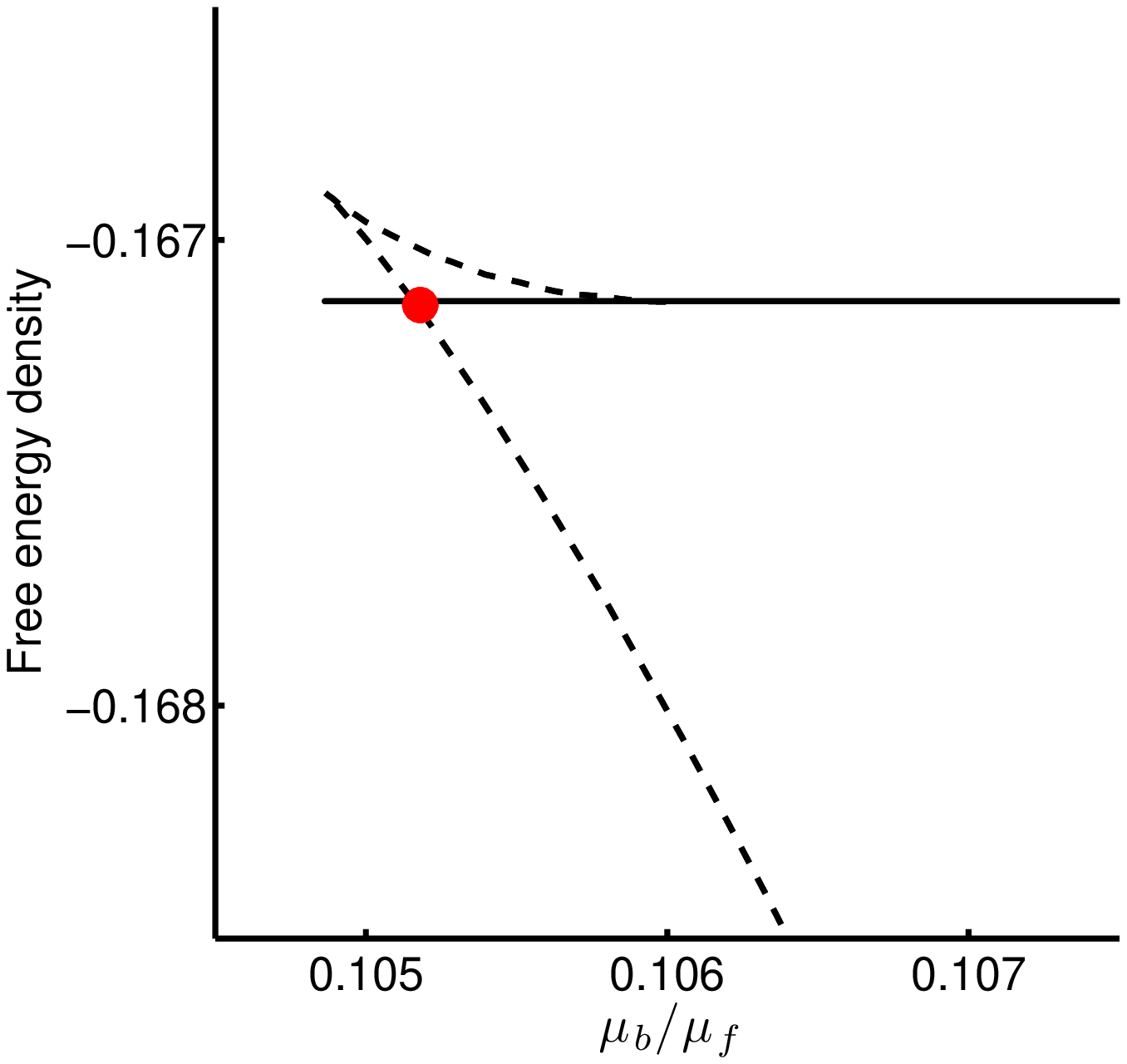}}
  \caption{(Color online) (a) Phase diagram of the Bose-Fermi
    superfluid mixture at $T=0$ \cite{Numbers}. Solid (green) curve is
    the dynamical stability contour while the dashed (blue) contour
    $C$ denotes points in phase space with a fixed value of
    $\lambda_f$ such that $T_c= 0.11\, T_f$. The filled (red) circle
    represents the critical point, for this specific experimental
    realization, at which the homogeneous mixture enters the
    dynamically and mechanically stable region. (b) Plot of the free
    energy densities of pure fermions (solid) and the homogeneous mixture
    (dashed) against $\mu_b[C]$. Free energy of homogeneous
    mixture is lower than that of pure fermions only beyond the
    critical point represented by the filled (red) circle. Here, free energy density is a dimensionless quantity \cite{freeenergy}.}
 \end{figure*}
 For illustrative purposes, we start with a brief discussion of the zero-temperature phase diagram. Throughout,
 we follow the scheme outlined in Sec.~\ref{derivefree} to construct
 all the phase diagrams.
\subsection{Zero-Temperature limit}
\label{secZeroT}
\par
This is simply derived by taking the $T=0$ limit of
Eqs.~(\ref{gap})-(\ref{Hall}).  In the phase diagram shown in
Fig.~\ref{pdT0}, the solid (green) curve represents the boundary of
the dynamically stable region above it, separating the unstable region
below. However, it is important to note that, in the phase diagram the
interaction parameter $\lambda_f$ is determined self-consistently from
the Gap equation. Thus in any single experimental realization, only a
small portion, corresponding to a fixed $\lambda_f$, of the above
phase space is accessible. In our analysis, we choose a value of
$\lambda_f$ (corresponding to $1/k_f \, a_f = -1.10$) such that pure
fermions are in the BCS superfluid regime. This corresponds to a BCS
superfluid critical temperature of $T_c= 0.11\, T_f$, with $T_f$ being
the Fermi temperature \cite{Overestimate}. It is however
important to note that $T_{c,\text{mix}}$, the
BCS transition temperature of fermions in the presence of bosons, is
modified by the presence of the factor $\lambda_{bf}\,n_b$ in the
effective fermion chemical potential. Further, the dependence is such
that $T_{c,\text{mix}} \leq T_c$ and the equality is satisfied when
$\lambda_{bf} n_b \rightarrow$ 0.
\par
Points in the phase space that correspond to this fixed value of
$\lambda_f$ is shown by the dashed (blue) contour $C$ in
Fig.~\ref{pdT0}. The crossing of this contour with the dynamical
stability contour indicates the phase space point at which the
homogeneous mixture enters the dynamically stable region. However, for
the homogeneous mixture to be the stable ground state, mechanical
stability condition should also be satisfied on top of the dynamical
stability condition. This additional criterion is exclusively present
due to the spatial inhomogeneity intrinsic in trapped-atom
setups. By mechanical stability, we mean that
the free energy of the homogeneous mixture should be less than the
free energy of the pure bosonic phase or the pure fermionic phase
along the contour $C$. The value of the latter is a constant, since
$\lambda_f$ is fixed along $C$. Free energy of pure bosons along $C$
is given by
\begin{equation}
f_b[C] = - \frac{\mu_b^2[C]}{2 \, \lambda_b}
\end{equation}
where, $\mu_b[C]$ is the boson chemical potential along $C$.  
\par
The plot of Fig.~\ref{freeT0} shows the comparison of free
energies mentioned above. The free energy of the pure fermionic phase
and that of the homogeneous Bose-Fermi mixture along $C$ are
represented by the solid and the dashed lines respectively. The
free energy of pure bosons is much higher than the others and hence
bosons do not phase separate out of the mixture. Actually this
observation is a general property of the phase space of a
3-dimensional Bose-Fermi mixture. The filled circle (red) represents
the critical point along contour $C$, at which the free energy of the
homogeneous mixture is lower than that of pure fermions, i.e., the
critical point at which the homogeneous mixture enters a region of
both dynamical and mechanical stability. This implies that, for the
specific experimental realization considered here, up until this
critical point pure fermions phase separate out of the mixture, while
above the critical point the Bose-Fermi superfluid exists as a stable
mixed phase.
\subsection{Finite-Temperature scenario}
\label{secFiniteT}
 We begin the discussion of the finite temperature case by first
  emphasizing some of the generic aspects of such a phase diagram as
  depicted in the schematic of Fig.~\ref{cartoon}.
\subsubsection{Generic features}
\begin{figure}[ht]
\centering
\includegraphics[scale=0.5]{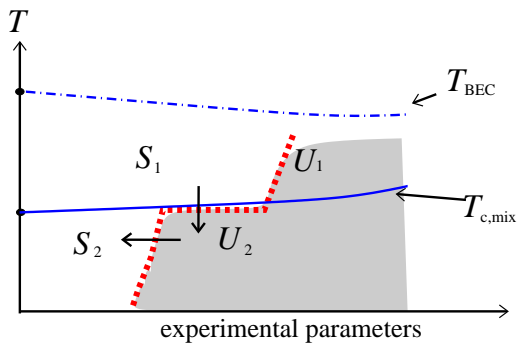}
\caption{(Color online) Schematic depicting possible first-order transitions occurring in Bose-Fermi mixtures across $T_{c,\textrm{mix}}$. Horizontal axis denotes the $\mathbb{R}^5$ phase space of experimental parameters defined by $\{\lambda_{b}, \lambda_{bf}, \mu_{f}, n_b, \Delta\}$.}
\label{cartoon}
\end{figure}
Similar to the above illustration of the zero-temperature limit, we analyze the stability of the
superfluid Bose-Fermi mixture in the vicinity of BCS critical
temperature for a wide range of temperature and other parameter
values. As mentioned earlier in Sec.~\ref{derivefree}, the phase space
is huge (5-dimensional) allowing for complicated boundaries between
stable ($S_i$) and unstable ($U_i$) regions of the homogeneous
mixture. Before proceeding to the detailed quantitative finite
temperature phase diagram in Fig.~\ref{PD}, we therefore summarize our
findings by pointing out the broad features, as depicted in
Fig.~\ref{cartoon}. In a certain projected subspace, the homogeneous
Bose-Fermi mixture becomes dynamically and/or mechanically unstable
towards phase separation through a first-order transition
$S_1\rightarrow U_2$, when cooled across $T_{c,\textrm{mix}}$. The
tunability of experimental parameters further allows us to access the
$U_2\rightarrow S_2$ transition at some fixed temperature below
$T_{c,\textrm{mix}}$. We also observe the existence of a parameter
regime where the homogeneous mixture remains unstable across
$T_{c,\textrm{mix}}$ going from $U_1 \rightarrow U_2$. If fermions
phase separate out of the unstable regions, then along the phase space
boundary between $U_1$ and $U_2$, and that between $U_2$ and $S_1$,
$T_{c,\textrm{mix}}$ is essentially $T _c$. Thus, in short, the
Bose-Fermi mixture exhibits rich mixing-demixing physics in the
vicinity of the BCS critical temperature. Particularly interesting is
the parameter regime exhibiting the first-order transitions
$S_1\rightarrow U_2 \rightarrow S_2$, which shows how the already
condensed bosons affect the nucleation of fermions when cooled across
the critical temperature, thereby clearly indicating direct
implication for the observation of Fermi superfluidity in trapped
mixtures. We therefore address this part of the phase space in more
detail.
\subsubsection{Quantitative features}
\begin{figure}[ht]
\centering
\includegraphics[scale=0.55]{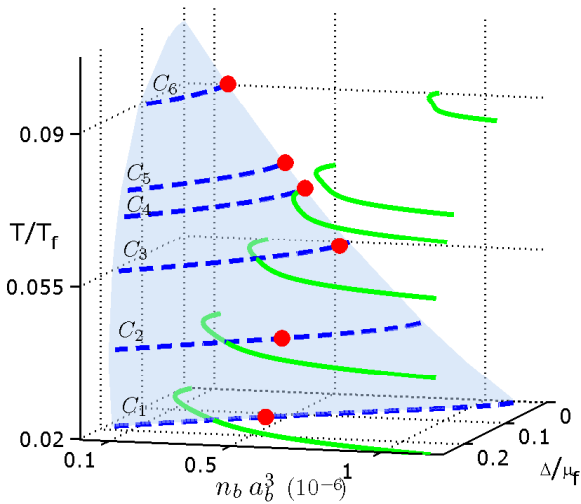}
\caption{(Color online) Phase diagram of the Bose-Fermi superfluid
  mixture at temperatures $T_c>T$. Solid (green) curves show dynamical
  stability criteria and the (blue) surface represents points of fixed
  $\lambda_f$, such that $T_c$ = 0.11 $T_f$ \cite{Overestimate, Numbers}. Dashed (blue) curves $C_i$
  are contours of fixed temperature $T_i$ along this surface and (red)
  circles indicate critical points at which the homogeneous mixture enters the region of both dynamical and mechanical stability.}
\label{PD}
\end{figure}
For temperatures $T>T_{c,\textrm{mix}}$ before the onset of BCS
superfluidity, $\Delta=0$ and hence we are confined to the $n_b$
axis. Correspondingly, the free energy and the stability conditions of
the homogeneous Bose-Fermi mixture are given by simply substituting
$\Delta = 0$ in Eqs.~(\ref{gap})-(\ref{Hall}). For the parameter
space under investigation, we find that the homogeneous mixture is
always the stable ground state in this temperature regime. On the
other hand, at temperatures $T<T_{c,\textrm{mix}}$, the onset of BCS
superfluidity in fermions is characterized by a non-zero value of
$\Delta$. In Fig.~\ref{PD}, we plot the phase diagram for a wide range
of temperatures below $T_c$ to observe that the mixture is dynamically
stable only above the solid (green) curves at a given
temperature. Thus the presence of an all-stable homogeneous phase
above $T_{c,\textrm{mix}}$ and a mixture of unstable/stable phases
below $T_{c,\textrm{mix}}$, as seen in Fig.~\ref{PD}, depicts the
unambiguous manifestation of $S_1\rightarrow U_2\rightarrow S_2$
transitions.
\par
Now we can immediately recognize the significance of this phase
diagram for a realistic experimental situation. Again, just like
  the $T=0$ case, only a small
part of the stable phase space, corresponding to a fixed value of
$\lambda_f$, is accessible in a particular experimental
realization. This we indicate by the two-dimensional surface shown in
Fig.~\ref{PD}, for our chosen value of $\lambda_f$ such that the
fermions are in the BCS superfluid regime. The dashed lines $C_i$'s
are contours connecting phase space points on this surface with fixed
temperatures $T_i$'s.  The crossing of contours $C_i$'s with dynamical
stability contours indicates the phase space points at which the
mixture enters the dynamically stable region. The filled circles
represent critical points at which the homogeneous Bose-Fermi
superfluid mixture becomes the stable ground state, i.e., both
dynamically and mechanically stable. We observe their occurrence to
transpire in two different ways: (1) In $C_1$-$C_4$, critical points
occur in the dynamically stable region where the mixture also attains
mechanical stability (as illustrated in Sec.~\ref{secZeroT}). (2)
Along $C_5$ ($C_6$), the BCS transition temperature monotonously
reduces with $\Delta$ by such an extent that when $\Delta \rightarrow
0$, $T_{c,\textrm{mix}} < T_5$ ($T_6$). However as discussed before,
the mixture is always the stable ground state for
$T_{c,\textrm{mix}}<T$. Hence in $C_5$-$C_6$, the critical points are
given by their intersections with $\Delta$=0 plane. Thus along each
$C_i$, below these critical points the homogeneous mixture becomes
dynamically and/or mechanically unstable. We further find that in the
unstable regions, pure fermions phase separate from the mixture. Thus
in Fig.~\ref{PD} we clearly demonstrate the occurrence of
temperature-driven mixing-demixing transitions at fixed interaction
strengths.
\section{Trap Profiles within LDA}
\label{secLDA}
 \begin{figure}[ht]
\includegraphics[scale=0.5]{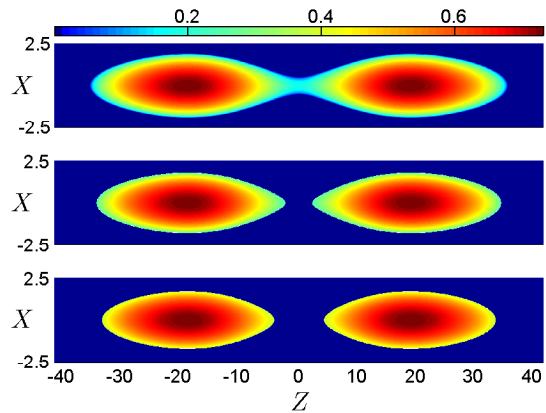}
\caption{(Color online) Subplots top, middle and bottom show boson
  density profiles (slice along y=0 plane) computed within the LDA for
  temperatures $T=\,$0.13, 0.072, 0.055 $T_f$ respectively, with $T_c$=0.11
  $T_f$. $\mu_b$ is adjusted to ensure
 number conservation ($\sim$42000 atoms).  $X$, $Z$ are in $\mu$m. Color bar shows density variations in scale of $n_ba_b^3$ (10$^{-6}$). Chosen trap parameters $V_0$ (several mW), $\sigma$=12\,$\mu$m and $\omega_{z}$/$\omega_{\perp}$=0.08.}
  \label{DP}
\end{figure}
We now show the direct experimental implications of the
above phase stability analysis. This, we do by reliably translating
this analysis to the inhomogeneous case via Local Density
Approximation (LDA) by defining a position dependent chemical
potential $\mu_b({\bf r})=\mu_b-V_{\text{trap}}({\bf r})$, where
$\mu_b({\bf r})$ is the local chemical potential and
$V_{\text{trap}}({\bf r})$ is the trap potential for bosons
\cite{reviewBEC}.  While this approximation is known to be very
efficient for large densities (typically the case in trapped-atom
experiments), it also implies that the trap potential can
simultaneously accommodate one or more of the phases discussed
above. Thus, remarkably what phase/phases will be observed will
critically depend on the trap geometry. This in fact, is a very
important observation implying the possibility of tuning the trap
parameters such that a desired density profile is observed only if a
certain phase has nucleated.
\par
To illustrate this program, let us consider bosons to be in a tightly
confined trap surrounded by the Fermi gas in a larger trap, a scenario
that takes advantage of our framework to consider a homogeneous Fermi
gas with fixed $\mu_f$. Additionally, this consideration is completely
justified as the trapping potential for each species can be
independently controlled \cite{KetterleSympathetic}. After careful
analysis of the phase diagram in Fig.~\ref{PD}, we find it
advantageous to confine bosons in a trap with a finite barrier near
the center. As this also helps to enhance the contrast in imaging the
nucleated phases, we propose a double-well cigar shaped trap with a
potential
\begin{equation*}
 V_\text{trap}(\textbf{r}) =  \frac{1}{2} m \omega_{\perp} (x^2+y^2) + \frac{1}{2} m \omega_{z} z^2 + V_0 \, \text{exp}(-\frac{z^2}{2 \sigma^2})
\label{Vtrap}
\end{equation*}
to confine bosons, where $\omega_{\perp}$ ($\omega_{z}$) is the
  trap frequency in the transverse (longitudinal) direction to the
  Gaussian beam creating the trapping potential. $V_0$ and $\sigma$,
defining the barrier peak and beam-width respectively, are chosen to
ensure a readily detectable overlap of boson density profiles from the
two wells for $T=0.13 \, T_f$ (i.e., $T_c<T$), as shown in the top
plot of Fig.~\ref{DP}. At $T=0.072 \,T_f$ ($0.055 \,T_f$), phase
stability analysis along contour $C_5$ ($C_3$) in Fig.~\ref{PD}
indicates the existence of a critical boson density (and
correspondingly a critical boson chemical potential $\mu_b(r)$), only
above which the Bose-Fermi mixture homogeneously co-exists as the
stable ground state. Corresponding regions of the trap where this
condition is not satisfied are devoid of bosons in a drastic fashion,
as seen from the $\sim$4 $\mu$m (8 $\mu$m) gap between the separated
bosonic islands in Fig.~\ref{DP}. As these separation lengths are far
greater than the healing length of the condensate, this illustration
vividly shows how crucial aspects of the finite temperature phase
diagram readily translate into detectable signatures in
experiments. Furthermore, this particular signature in Fig.~\ref{DP}
may be used as a signal indicating the onset of BCS superfluidity in the
  particular parameter regime of the attractive Fermi gas.
\section{Effect of Dimensionality on the Phase diagram}
\label{sec1D}
As a final piece, we analyze how the phase diagram gets modified when
only the trap geometry of the experimental setup is deformed (all
other parameters kept constant) such that the confinement in two
orthogonal directions is made much tighter compared to that in the
third. In effect, the system can be considered to
be one-dimensional if the trapping frequency in the tight
directions is such that $\hbar \, \omega_{\text{tight}} \gg
\mu_{b,f}$. For simplicity, we restrict ourselves to the $T=0$ limit,
where all the qualitative features can be comprehensively discussed
\cite{Satyan1D}. The effective 1-d interaction strength can be written
in terms of the 1-d scattering length, which in turn can be easily
related to the 3-d scattering length \cite{Olshanii}. This mapping is
critically dependent on the aspect ratio of the trap. We choose
experimentally relevant values for $\omega_\perp$ and $\omega_z$
\cite{Hulet1D}. While we assume $\omega_\perp \simeq 2 \, \pi \, 10^5$
Hz for both bosons and fermions, we find it useful to consider
$\omega_\perp / \omega_z \simeq \, 10^3$ for fermions but smaller
values of $\omega_\perp / \omega_z$ for bosons. Apart from ensuring
that we are indeed in the 1-d regime, this choice guarantees a highly
elongated trapping potential for fermions. From the parameter values
used in deriving the 3-d phase diagram of Fig.~\ref{pdT0}, we obtain
the corresponding values for the 1-d scenario \cite{Olshanii}.

\begin{figure*}[t]
  \centering
  \subfigure{\label{pdT01D}\includegraphics[width=2.75in,height=2.5in,clip]{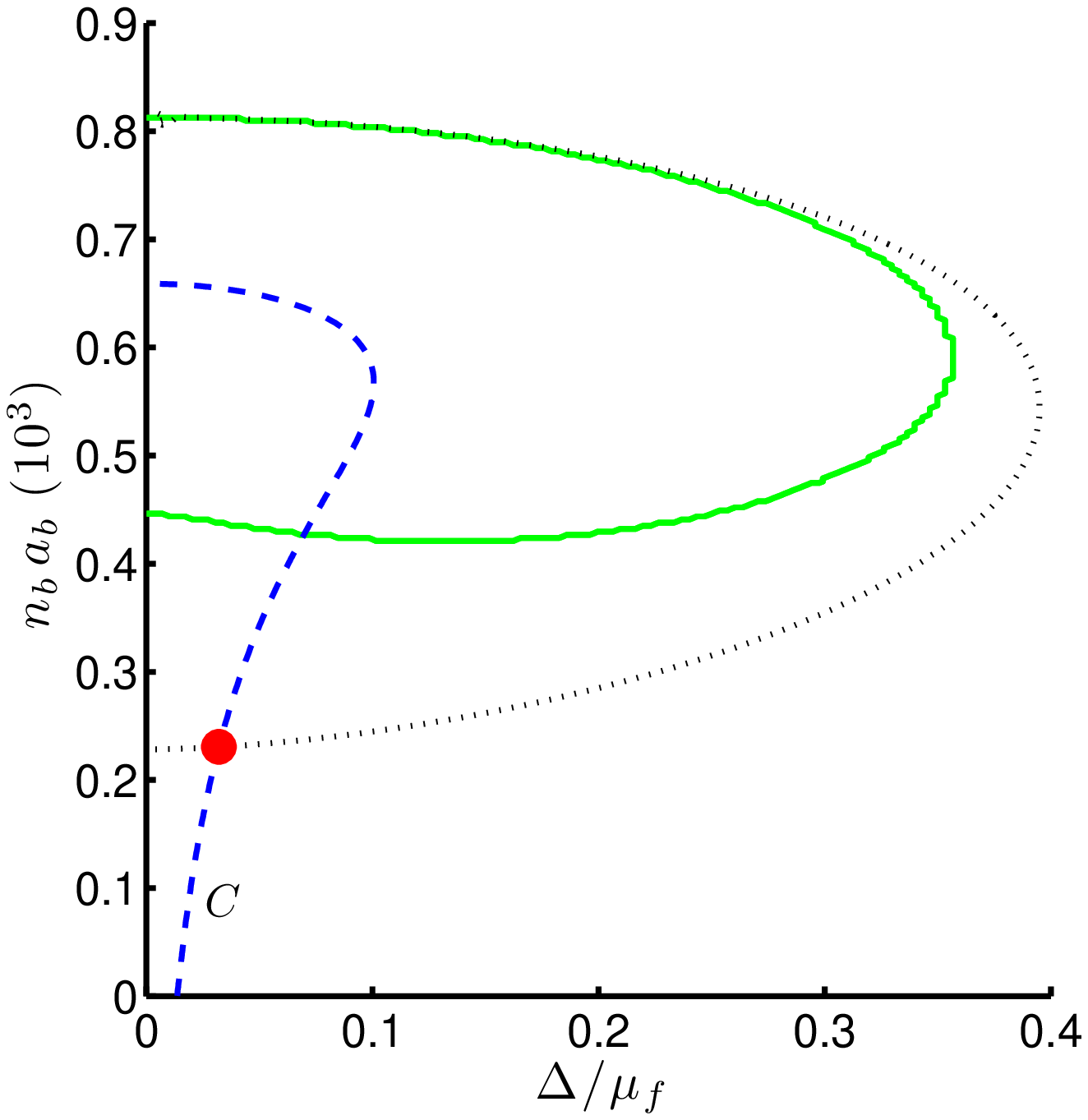}}
  \subfigure{\label{freeT01D}\includegraphics[width=2.75in,height=2.5in,clip]{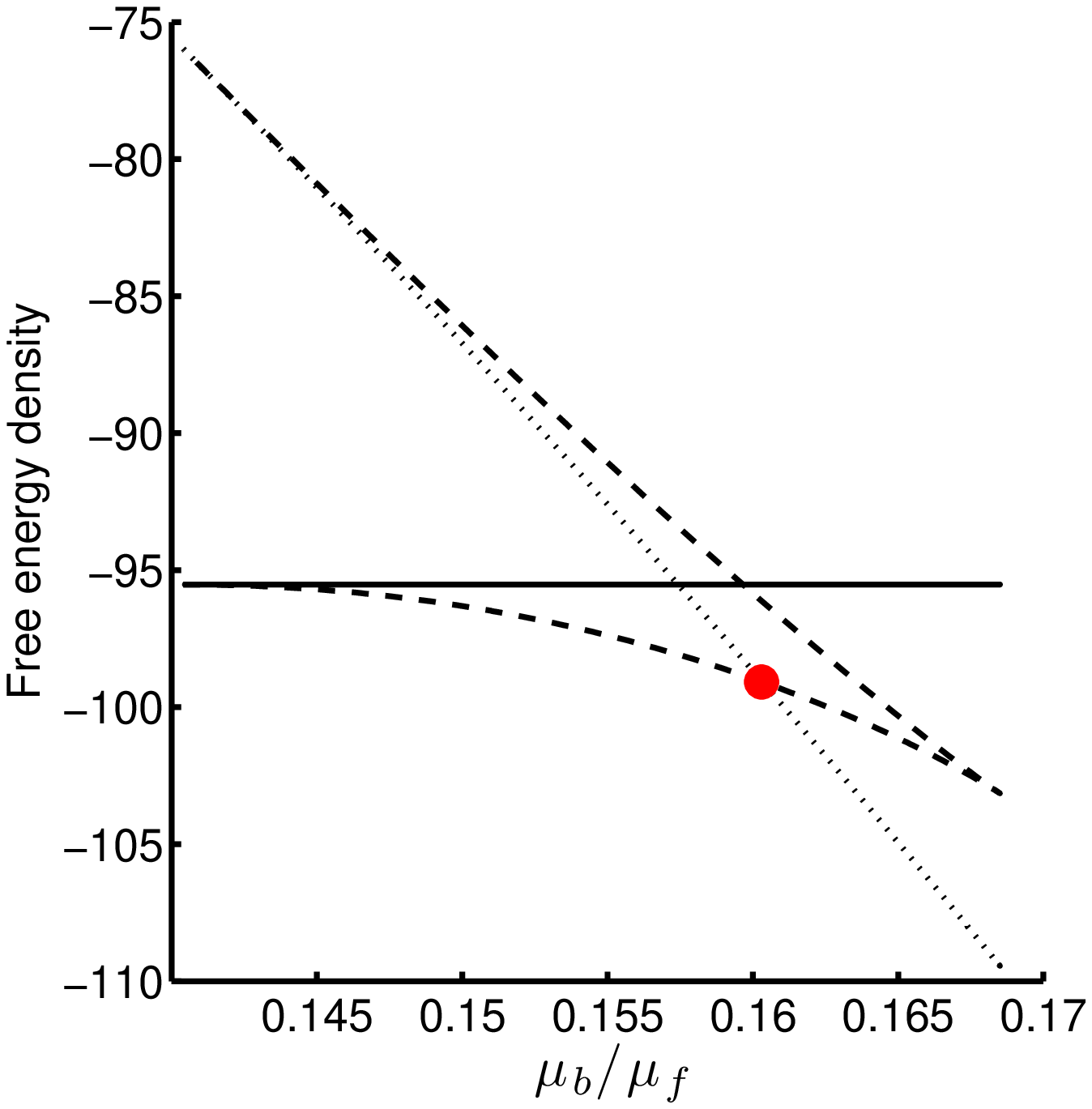}}
\caption{(Color online) (a) Phase diagram of the Bose-Fermi superfluid
  mixture in 1-d at $T= 0$ (for the same parameters used in
  Fig.~\ref{pdT0}). Solid (green) curve is the dynamical
  stability contour, while the dashed (blue) contour $C$ denotes phase
  space points with a fixed value of $\lambda_{1D\,f}$. The dotted (black) curve represents the
  region within which the homogeneous mixture is mechanically
  unstable. The filled circle (red) represents the critical point, for
  this specific experimental realization, at which the homogeneous
  mixture enters the mechanically unstable region. 
  (b) Plot of the free energy densities of pure bosons
  (dotted), of pure fermions (solid) and the
  homogeneous mixture (dashed) against $\mu_b[C]$. Free
  energy density of homogeneous mixture becomes higher than that of pure
  bosons at the critical point represented by the filled (red) circle,
  resulting in the phase separation of pure bosons out of the
  mixture. Here, free energy density is a dimensionless quantity \cite{freeenergy}.}
 \end{figure*}

 We thus construct the phase diagram of the
 interacting one dimensional superfluid mixture, by performing the 1-d
 integrals instead of 3-d in Eqs.~(\ref{gap})-(\ref{Hall}).
 The phase diagram is shown in Fig.~\ref{pdT01D}, where the
 solid (green) curve represents the dynamical stability contour that
 separates the dynamically unstable region (inside the ellipse) from
 the dynamically stable region (outside the ellipse). Phase space
 points that correspond to the fixed value of $\lambda_{1d\,f}$,
 are shown by the dashed (blue) contour $C$. The crossing of $C$
 and the dynamical stability contours indicates the point at which the
 homogeneous mixture enters the dynamically
 unstable region. However, as discussed before in Sec.~\ref{secZeroT},
 for the Bose-Fermi homogeneous mixture to be stable, it is necessary that the mechanical
 stability condition be simultaneously
 satisfied. For this, we plot the relevant free energies
 in Fig.~\ref{freeT01D}, where the free energies of pure bosons, homogeneous mixture and pure
 fermions along $C$ are given by dotted, dashed and solid lines respectively.
  We immediately note that pure fermions can never phase separate out
 of the mixture, a remarkably different result
 when compared to the 3-d case [see Fig.~\ref{freeT0}]. The filled circle
 (red) represents the critical point along contour $C$, at which the
 free energy of the bosons becomes lower than that of the homogeneous
 mixture, i.e., the critical point at which the homogeneous mixture
 becomes mechanically unstable. This means that up until this critical
 point, homogeneous superfluid mixture coexists as the stable ground
 state. However, above this critical point pure bosons phase separate
 out of the mixture. The significance of this critical point is clear
 since we can now directly obtain the boson density profile in the
 boson trap by mapping the boson chemical potential $\mu_b[C]$ onto
 the spatial coordinate in the trap via LDA using $\mu_b[C] = \mu_b(r)
 = \mu_b - V(r[C])$, where $V(r)$ is the probe trapping potential for
 the bosons. It is evident from Figs.~\ref{pdT01D} and ~\ref{freeT01D}
 that the pure bosons phase separate out of the homogeneous mixture in
 the trap above the critical value of boson chemical potential $\mu_b$
 corresponding to the critical point (filled red circle).

\section{Conclusions}
\label{secConclusion}
\par
In this article, we have discussed a consistent
theoretical method for performing the finite
temperature phase stability analysis of an ultra-cold mixture
comprising of bosons and fermions, both in the superfluid
regime. Based on our stability analysis in the vicinity of the
  Fermi superfluid temperature, we discussed two distinct scenarios where the
homogeneous superfluid mixture becomes unstable (1) when the
normal-superfluid phase transition (second-order) occurs in the fermionic component, and 
(2) below the Fermi superfluid temperature via mechanical instability
which is a first-order phase-separation phase transition. The latter
scenario happens exclusively due to the trap inhomogeneity inherent in
trapped-atom experiments, thereby allowing for the two phases to be
simultaneously present. We have illustrated the emergence of these
  instabilities and the ensuing phase separation by considering a
realistic experimental setting. We fine-tuned the trap geometry to
enhance the effect of phase separation. Finally, we briefly discussed
the effect of dimensionality on the stability of various phases. We reiterate that while our phase diagram analysis is quantitatively exact when the interactions are weak, our study is only qualitatively correct in the strong interaction limit.
\par
We emphasize here that the interplay between the first and second
order phase transitions, similar to that discussed in this paper, will
have strong implications for analyzing experimental observations
involving ultra-cold mixtures in general \cite{Leslie}. While our 
framework is also valid to study the regime of strong interactions
near a broad Feshbach resonance, it can be easily extended within a
two channel model for the case of a narrow Feshbach
resonance. Further, while such a treatment will naturally allow for a
molecular condensate of Fermi atoms \cite{Jin}, it is not hard to
speculate emergence of rich physics due to occurrence of Efimov bound
states in the Bose-molecule interaction channel
\cite{Efimov}. Finally, an important extension of the current work
would be to consider spin-dependent Bose-Fermi interactions. The
presence of a small BEC can shift the chemical potential of a
particular spin component relative to the other.  This is analogous to
the situation encountered in solid-state samples with magnetic
impurities, thereby providing a new platform for studying the
interplay between superfluidity and magnetism.
 \\
  \section*{Acknowledgments}
  This work is supported by the W. M. Keck Program in Quantum
  Materials at Rice University, NSF, and the Robert A. Welch
  Foundation (Grant No. C-1669). S.B. also acknowledges funding
  from the Centre for Quantum Science at George Mason University.


\begin{thebibliography}{99}
\bibitem{Helium} Experimental realization has not been possible due to extremely low superfluid transition temperature of $^3$He.
\bibitem{Altland1} Even if $\Delta$ is not real and $\Delta = |\Delta| e^{i \phi}$, it can always be made real by the global gauge transformation \cite{Altland}.
\bibitem{muf} This choice is especially useful if we consider the bosons to be in a tight trap enclosed by the fermions in a larger trap.
\bibitem{diverges} In 3-d, the term $\sum_{k} (\xi_k - E_k)$ itself diverges and hence it is useful to add the term $\sum_k \Delta^2 /2 \, \varepsilon_k$. At equilibrium, the absolute value of free energy $f$ remains unchanged due to regularization, since we have simply included terms $+ \sum_k \Delta^2 /2 \, \varepsilon_k$ in $\sum_{k} (\xi_k - E_k)$ and $- \sum_k \Delta^2 /2 \, \varepsilon_k$ in $\Delta^2/|\lambda_{f}|$.
\bibitem{Numbers} Here, chosen values are: $T_f$=$\mu_f/k_B$=3.8 $\mu$K ($m$=6 $amu$), $a_b$=35 $a_0$ and $a_{bf}$=230 $a_0$, where $a_0$ is the Bohr radius and $\mu_f$ is fixed in our analysis.
\bibitem{freeenergy} Free energy density is made dimensionless by scaling with respect to the energy density of a 3-d non-interacting gas occupying a volume of $1/k_f^3$ with fermi-momentum $k_f = 1/3200 \, a_0$ and $m=6 \, amu$. Corresponding vales in the 1-d limit are computed using the results in \cite{Olshanii}.
\bibitem{Satyan1D} Phase diagram of the Bose-Fermi superfluid mixture in 1-d at $T=0$ was already discussed by two authors of this manuscript in Ref.~\cite{Satyan1Dp}.
\end{thebibliography}
\end{document}